\documentclass[useAMS,usenatbib]{mnras}

\usepackage[usenames,dvipsnames]{xcolor}

\usepackage{aas_macros}
\usepackage{amssymb,amsmath}
\usepackage{natbib}
\usepackage{graphicx}

\newcommand{\teffmax}{T_\mathrm{eff,\,max}^\infty}
\newcommand{\tmax}{T_\mathrm{max}^\infty}

\title[R-modes and neutron star recycling scenario]
{R-modes and neutron star recycling scenario}

\author[A. I. Chugunov, M. E. Gusakov, E.M. Kantor]
{A.~I.~Chugunov$^1$,\thanks{E-mail: andr.astro@mail.ioffe.ru},
 M. E. Gusakov$^{1}$,
 E. M. Kantor$^1$\\
 $^1$Ioffe Institute, St Petersburg, Russia
\\
}

\begin{document}

\date{Accepted 2017 xxxx. Received 2017 xxxx;
in original form 2016 xxxx}

\pagerange{\pageref{firstpage}--\pageref{lastpage}}
\pubyear{2017}

\maketitle

\label{firstpage}

\begin{abstract}
To put new constraints on the r-mode instability window, we
analyse the formation of millisecond pulsars (MSPs) within the
recycling scenario,
making use of
three sets of observations: (a) X-ray observations of
neutron stars (NSs) in low-mass X-ray binaries; (b) timing
of millisecond pulsars; and (c) X-ray and UV
observations of MSPs. As shown in previous works, r-mode
dissipation by shear viscosity is not sufficient to explain
observational set (a), and enhanced r-mode dissipation at
{the red-shifted} internal temperatures $T^\infty\sim
10^8$~K is required to stabilize the observed NSs. Here, we
argue that models with enhanced bulk viscosity can hardly
lead to a self-consistent explanation of observational set
(a) due to strong neutrino emission, which is typical for
these models (unrealistically powerful energy source is
required to keep NSs at the observed temperatures). We also
demonstrate that
the observational set (b)
{, combined with the theory of internal heating and NS
cooling, provides evidence of}
enhanced
r-mode dissipation at low temperatures,
{$T^\infty\sim 2\times 10^7$~K}. Observational set (c)
allows us to set an upper limit on the
 internal
temperatures of MSPs, $T^\infty<2\times 10^7$~K (assuming a
canonical NS with the accreted crust). Recycling scenario
can produce MSPs at these temperatures only if r-mode
instability is suppressed in the whole MSP spin frequency
range ($\nu\lesssim 750$~Hz) at temperatures $2\times
10^7\lesssim T^\infty\lesssim
3
\times 10^7$~K, providing thus a new constraint on the
r-mode instability window. These observational constraints
are analysed in more details in application to the
resonance uplift scenario of \cite{gck14a,gck14b}.
\end{abstract}

\section{Introduction}

According to the generally accepted recycling scenario
{(\citealt{bkk76, acrs82})}, millisecond pulsars (MSPs) are
descendants of neutron stars (NSs), which were spun up by
accretion from Roche-lobe filling companion stars in
low-mass X-ray binaries (LMXBs). This scenario has been
extensively studied in the literature with the main
emphasis on the NS spin frequency, evolution of
the companion star, and parameters of the binary system
(e.g., \citealt{Tauris12,ccth13} and references therein).
NS temperature has not been typically considered as an
important parameter for NS recycling, and  is not
even mentioned in most of the papers in the field. However,
NS temperature is crucial for the instability of r-modes
(similar to Rossby waves controlled by the Coriolis force)
driven by Chandrasekhar-Friedman-Schutz (CFS;
\citealt{chandrasekhar70a,fs78a,fs78b}) mechanism due to
emission of gravitational waves
(\citealt{andersson98,fm98}).
Namely, if a rapidly rotating NS is hot enough, it becomes
unstable with respect to excitation of r-modes (see, e.g.,
\citealt*{lom98}; the corresponding region of NS
temperatures and spin frequencies is often referred to as
`the instability window'). This instability is known to
modify the NS evolution in LMXBs dramatically (see section
\ref{Sec_minmod} and, e.g.,
\citealt*{levin99,btw07,gck14b}). In particular, it can
limit rotation frequencies of NSs
(\citealt*{bildsten98,aks99}).
Furthermore, we (\citealt*{cgk14})
argued that the r-mode instability can lead to additional
channel of NS recycling:
formation of a new class of
NSs
--- HOFNARs
(from HOt and Fast Non-Accreting Rotators)
--- in addition to MSPs.
Thus, accurate account for the r-mode instability should be
important for the recycling scenario; conversely,
verification of the recycling scenario can put stringent
constraints on the r-mode instability and, therefore, on
the properties of superdense matter.

In this paper we analyse constraints on the r-mode
instability within the recycling scenario. We base our
analysis on the following observations:
\begin{itemize}
\item[(a)] X-ray observations of
transiently accreting NSs in LMXBs (section
\ref{Sec_Xray});
\item[(b)] Timing of MSPs (section \ref{Sec_timing});
\item[(c)]
X-ray and \textit{UV} observations of MSPs (section
\ref{Sec_Xray_MSP}).
\end{itemize}
We do not take into consideration observational data on
binary system parameters (orbital period, eccentricity
etc.), because they seem to be determined by the evolution
of binary system (e.g.,\ \citealt{ccth13}) and not affected
directly by the r-mode instability.

The set of observations (a) allows one to estimate the
{effective} surface temperatures (in the quiescent state,
see, e.g., \citealt{hjwt07,heinke_et_al_09}) and spin
frequencies (e.g., \citealt{watts_et_al_08}) for a number
of accreting NSs. It has been shown to be a crucial test of
the r-mode instability theory (\citealt*{hah11,hdh12}). In
particular, those references have demonstrated that the
`minimal' model of r-mode instability window, suggested by
\cite{lom98} (NS has a nucleonic core, r-mode damping is
associated with shear and bulk viscosities, see section
\ref{Sec_minmod}), should be supplemented by additional
damping mechanism at
 redshifted internal temperatures
$T^\infty\sim (3\times 10^7-10^8)$~K%
\footnote{Following \cite{gck14a,gck14b}, we characterize
thermal states of NSs by the redshifted temperature
$T^\infty$, rather than by the local temperature $T$,
because it is $T^\infty$, which is constant throughout the
star. }
%
 (alternatively, saturation amplitude
for r-modes should be extremely small, the so called `tiny
r-mode amplitude scenario', see section \ref{Sec_tiny} and,
e.g., \citealt{ms13}).
 A number of papers are devoted to the
identification of the required damping mechanism (see the
recent review by \citealt{haskell15}). In particular,
\cite{gck14a,gck14b} argues that finite temperature effects
in superfluid NS cores can lead to resonance mode coupling
and suppression of the r-mode instability at certain
stellar temperatures.
As a result, the instability window
is splitted up by `stability peaks' in the vicinity of
these temperatures. \cite{gck14a,gck14b} discuss evolution
of NSs in LMXBs within this model and suggest a `resonance
uplift scenario', which allows
them
to explain the observational set (a). Other ideas about the
required additional damping include enhanced mutual
friction (e.g., \citealt{hap09}), crust-core coupling
(e.g., \citealt{Rieutord01,lu01,ga06a,ga06b}), and exotic
core composition (e.g.,
\citealt{Jones01_comment,lo02,no06,as14_msp}). Concerning
the observational set (a) we point out that any successful
theory should not only stabilize the observed NSs in LMXBs,
but also explain their temperatures.

MSP timing [set of observations (b); see section
\ref{Sec_timing}] has also been used to constrain the
r-mode instability.
\cite{rb03} discuss formation of MSPs assuming bulk
viscosity driven suppression of r-mode instability. They
predict that spin down of the fastest MSPs can be affected
by r-modes. Furthermore, these NSs should have internal
temperatures $T^\infty \sim 2\times 10^7$~K, being the
sources of thermal X-rays.
 \cite{gck14b} analyse formation of
recycled NSs within the resonance uplift scenario and
concluded that it does not contradict the idea that MSPs
are descendants of NSs in LMXBs. They also predict that the
most rapidly rotating MSPs should be rather hot (with
internal temperatures $T^\infty\sim 10^7$~K) due to r-mode
heating. \cite{owen10,as15_oscMSP} argue that the minimal
instability model requires extremely small r-mode
saturation amplitudes to match the observed pulsars'
spin-down.
\cite{as14_msp} discuss MSP formation and
conclude that compact stars with the cores composed of
`ungapped interacting quark matter' are consistent with
observations of LMXBs and MSPs (see a critique of this
statement in section \ref{Sec_bulk}). They also confirm
that additional damping is required to describe MSPs by the
models with nucleonic core composition. In this paper
(section \ref{Sec_timing}) we present an independent
argument in favour of enhanced r-mode damping at low
temperatures, based on the MSP internal heating mechanisms:
superfluid vortex creep (\citealt{Alpar_etal84}),
rotochemical heating (\citealt{Reisenegger95}), and
rotation-induced deep crustal heating (\citealt*{gkr15}).

The X-ray and \textit{UV} observations [set (c), section
\ref{Sec_Xray_MSP}] attract less attention in the r-mode
literature, because the data for most of the observed MSPs
do not require thermal emission from the whole surface (but
only from hot spots; see, e.g.,
\citealt{Zavlin07,Bogdanov_etal11_M28}) and the surface
temperature is measured only for PSR J0437-4715 by
\cite{Durant_etal12_0437} { and for PSR J2124$-$3358 by
\cite{Rangelov_etal17}}.
 For other MSPs the surface temperature is
thought of as
{`generally unknown'} 
(\citealt{as14_msp})
or not too high
 {[namely,
\citealt{rb03} argue that MSP temperatures are `below the
upper limits from observations of MSPs with the ROSAT
(Roentgen Satellite) X-ray telescope' and
\citealt{gck14b,cgk14} assume that redshifted surface
temperature of MSPs can be bounded by
$T^\infty_\mathrm{eff}\lesssim 10^6~K$].
Upper limits for internal temperature of several MPSs are also shown in Fig.\ 1 of \cite*{ajk02}.}
 In this paper we argue that X-ray
and \textit{UV} observations allow one to put an upper limit on the
redshifted surface temperature of MSPs,
$T^\infty_\mathrm{eff}< \teffmax$, and argue that
$\teffmax
=6\times10^5$~K can be chosen as a fiducial value,
applicable to all MSPs (otherwise thermal emission from the
whole surface should be observed, see section
\ref{Sec_Xray_MSP}). It corresponds to an upper limit on
the internal temperatures of MSPs: $T^\infty< \tmax=
2\times 10^7$~K (here and below, to estimate internal
temperatures of NSs, we consider a canonical NS with the
mass $M=1.4 M_\odot$ and $R=10$~km, and use the accreted
envelope model from \citealt{pcy97}). This constraint is
crucial for r-mode instability models, because any
successful model should not only stabilize NSs at these
temperatures (in the opposite case r-mode instability heats
MSPs up to higher temperatures), but also explain how they
came into this temperature region.

In sections  \ref{Sec_tiny}--\ref{Sec_gen} we apply
the observations (a--c) to constrain the r-mode instability
parameters. Namely, in section \ref{Sec_tiny} we improve
constraints on the tiny r-mode amplitude scenario. In
section \ref{Sec_bulk} we analyse models based on an
enhanced bulk viscosity, in particular, the ungapped
interacting quark matter model, suggested by
\cite{as14_msp}, and conclude that it
{cannot} explain temperatures of the hottest NSs in LMXBs.
In section \ref{Sec_reson} we analyse the resonance uplift
scenario. We demonstrate that low-temperature stability
peaks at $T^\infty\lesssim 2\times 10^7$~K are required by
observations, and
{they should} be broad. These results are extended to a
quite general class of r-mode instability models
{(a set of humps of enhanced stability located at certain
temperatures)}
in section \ref{Sec_gen}. We conclude in section
\ref{Sec_Consl}.

During the final stage of the preparation of this
manuscript we became aware of the work by
\cite{Schwenzer_etal_Xray}, who also use X-ray observations
to constrain MSP temperatures and derive upper bounds on
the r-mode amplitudes for these sources. The results of our
section \ref{Sec_tiny} qualitatively agree with those of
\cite{Schwenzer_etal_Xray}.

The preliminary results of this work were presented by one
of us (AIC) at  two conferences: (i) `The Modern Physics
of Compact Stars and Relativistic Gravity 2015' (September
30 -- October 3; Yerevan, Armenia, 2015); and (ii) `The
International Workshop on Quark Phase Transition in Compact
Objects and Multimessenger Astronomy: Neutrino Signals,
Supernovae and Gamma-Ray Bursts' (October 7 -- October 14;
Nizhnij Arkhyz \& Terskol, Russia, 2015).

\section{Minimal r-mode instability model vs observations:
additional dissipation is required } \label{Sec_observ}

In this section we describe the minimal r-mode instability
model and confront it with each of the three sets of
observations (a)--(c) from Section 1.

\begin{figure}
        \includegraphics[width=\columnwidth]{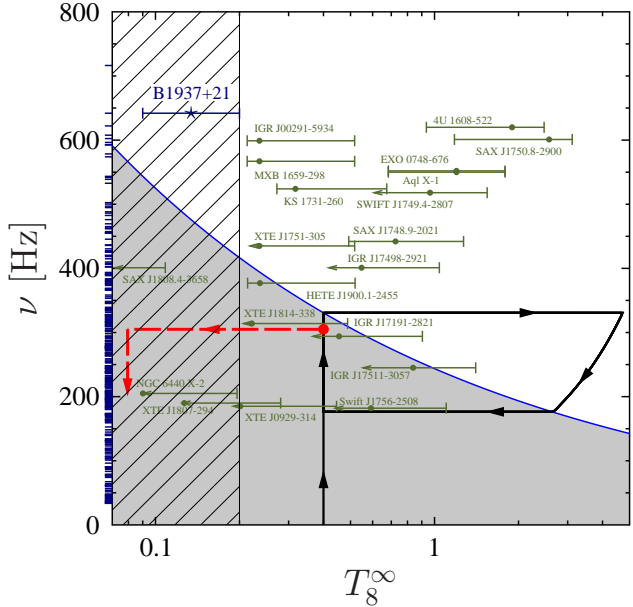}
   \caption{Minimal r-mode instability model for a canonical NS.
    The stability region is shaded in grey;
    in the white region the r-mode is unstable.
   Hatched region corresponds to the observational limit (\ref{Teffmax})
    on the MSP temperature.
    Temperatures and frequencies
    of NSs observed in LMXBs are shown by filled circles,
    error bars show uncertainties due to NS envelope composition (section \ref{Sec_Xray}).
    The pulsar B1937+21 is shown by star; its low-temperature error bar
    is set by the
    rotation-induced deep crustal heating (see section \ref{Sec_timing}),
    the upper error bar is set by the fiducial limit (\ref{Teffmax}) on the MSP temperature.
    Ticks on the left side of $\nu$ axis show measured MSP frequencies.
    Typical evolution track for a NS in LMXB is shown by a thick black line,
    and a large filled circle marks the end of LMXB stage.
    The dashed line demonstrates evolution track of a star in the MSP stage.}
    \label{Fig_MinMod}
\end{figure}

\subsection{Minimal r-mode instability model}
\label{Sec_minmod}

In the absence of damping the gravitational radiation
reaction force leads to exponential growth of the r-mode
amplitude on a timescale (see, e.g.,
\citealt*{lom98,fll16})
\begin{equation}
 \tau_\mathrm{GR}\approx-3000\, \mathrm{s}\
 M_{1.4}^{-1}\,
 R_{10}^{-4}\,
 \nu_{500}^{-6}. \label{tau_GW}
\end{equation}
Here NS mass $M$ and radius $R$ are normalized to the mass
and radius of the canonical NS: $M_{1.4}=M/(1.4\,
M_\odot)$, $R_{10}=R/(10\,\mathrm{km})$, and
$\nu_{500}=\nu/(500\,\mathrm{Hz})$. Here $\nu=\Omega/2\pi$
is the NS spin rate.
{The numerical coefficients in equation (\ref{tau_GW}) and
other equations in this section correspond to a Newtonian
star with polytropic equation of state ($P\propto \rho^2$).
}

The minimal r-mode instability model suggested by \cite{lom98}
includes shear and bulk viscosities as damping agents and
assumes nucleonic equation of state.
The shear viscosity damping time can be estimated as%
%
\footnote{ See a discussion of the electron and nucleon
contribution to the shear viscosity in \cite{gck14b}. Note
that the shear viscosity employed by \cite{as14_msp}
differs from that in \cite{gck14b} because the latter
authors assumed that protons are superconducting, while in
\cite{as14_msp} they were considered normal [see
\cite{sy08} for a detailed analysis of the
superconductivity effects on the shear viscosity]. }
%
%
\begin{equation}
  \tau_\mathrm{s}=2.2\times 10^5\,\mathrm{s}\, R_{10}^5
     M_{1.4}^{-1} (T_8^\infty)^2,
     \label{tau_S}
\end{equation}
where $T_8^\infty=T^\infty/10^8~\mathrm{K}$. The bulk
viscosity damping time was calculated by \cite*{lmo99}:
\begin{equation}
\tau_\mathrm{b}=2.8\times 10^{18}\,\mathrm{s}\, M_{1.4}
\nu_{500}^{-2} R_{10}^{-3} (T^\infty_8)^{-6}. \label{tau_B}
\end{equation}
The total damping time is
$\tau_\mathrm{d}=1/(\tau_\mathrm{s}^{-1}+\tau_\mathrm{b}^{-1})$,
however, the effect of bulk viscosity is negligible for
temperatures of MSPs and NSs in LMXBs. The r-mode is stable
if $\tau_\mathrm{GR}^{-1}+\tau_\mathrm{d}^{-1}>0$ (grey
region in Fig.\ \ref{Fig_MinMod}) and unstable in the
opposite case
$\tau_\mathrm{GR}^{-1}+\tau_\mathrm{d}^{-1}<0$ (white
region in Fig.\ \ref{Fig_MinMod}). In the unstable region
the r-mode amplitude is growing until it reaches
saturation, where the nonlinear effects become important
{(see, e.g., \citealt{Arras_etal03,btw07,bw13}; \citealt*{hga14})}.

Gravitational waves emitted by r-modes reduce stellar
angular momentum $J$ at a rate
\begin{equation}
\dot J_\mathrm{GR}=\frac{3\tilde J M R^2 \Omega
\alpha^2}{\tau_\mathrm{GR}}. \label{JGR}
\end{equation}
The dissipation of r-mode leads to additional heating power
(e.g., \citealt{as14,gck14b})
\begin{equation}
 W^\mathrm{r}=\frac{\tilde J M R^2 \Omega^2
 \alpha^2}{\tau_\mathrm{eff}}, \label{HeatRate}
\end{equation}
where  $\tau_\mathrm{eff}=|\tau_\mathrm{GR}|$ for saturated
and $\tau_\mathrm{eff}=\tau_\mathrm{d}$ for unsaturated
r-mode, respectively, and $\tilde J\approx 0.016$.

General equations that describe evolution of a NS affected
by r-mode instability were formulated by \cite{lom98} (see
also \citealt{levin99,hl00,as14,gck14a}) and the typical
evolution track is shown by a thick black line in Fig.\
\ref{Fig_MinMod}. During the initial phase NS is spinned up
by accretion and its temperature
$T^\infty=T^\infty_\mathrm{eq}$ is determined by the
balance between the cooling [the corresponding emissivity
is $L_\mathrm{cool}(T^\infty)$] and deep crustal heating
$W^\mathrm{DCH}$ (\citealt*{bbr98}):
\begin{equation}
L_\mathrm{cool}(T^\infty_\mathrm{eq})=W^\mathrm{DCH}=K
c^2\dot M.
\label{Teq}
\end{equation}
Here the coefficient $K\sim 10^{-3}$ determines efficiency
of deep crustal heating. As shown by \cite{levin99}, when a
NS enters the instability region due to continuous
accretion, the r-mode amplitude starts to grow and its
subsequent evolution is very fast (provided that the
saturation amplitude is not too small%
{, see appendix B in \citealt{gck14b} for a set of
analytical estimates}). Eventually,  the star rapidly heats
up and slows down, leaving the instability region. In the
stable region it gradually cools down to the temperature
$T^\infty_\mathrm{eq}$, and a new spin up cycle starts
again. In principle, during its life a NS in LMXB can
complete several such cycles but it is important that the
probability to catch it in the instability region is
negligible.
After
Roche-lobe decoupling (see the filled circle in the evolution
curve in Fig.\ \ref{Fig_MinMod}) a star cools down and an
MSP is formed (MSP evolution track is shown by a dashed
line). MSP spin frequency is limited by the r-mode
instability curve (\citealt*{bildsten98,aks99b}).

\subsection{X-ray observations of
transiently accreting neutron stars in LMXBs}
\label{Sec_Xray}

X-ray observations of some of the transiently accreting NSs
allow one to measure their surface temperature in the
quiescent state (e.g., \citealt{hjwt07,heinke_et_al_09}).
The internal temperature can be inferred from the surface
temperature, assuming thermally relaxed envelope, with the
help of analytical fitting formulas presented in
\cite{pcy97}. Furthermore, the spin frequency can be
measured for accretion- and nuclear-powered MSPs (e.g.,
\citealt{watts_et_al_08}). As a result, both spin frequency
and internal temperature are available for
a rather large number of objects
 (\citealt{hah11,hdh12,gck14b}).

In addition, observations of transiently accreting NSs in
the active state allow one to estimate average accretion
rate and the corresponding heating power associated with
nuclear reactions in the accreted crust (e.g.,
\citealt{hjwt07,heinke_et_al_09}).

In this paper we employ the data for 20 NSs from  table
I of \cite{gck14b}, where a number of misprints found in
the previous works have been corrected (see dots with error
bars in Fig.\ \ref{Fig_MinMod}, temperature for HETE
J1900.1-2455 was updated according to
\cite{Degenaar_etal16_J1900},
$T^\infty_\mathrm{eff}=6.3\times10^5$\,K). At least 8 NSs
among them are predicted to be r-mode unstable within the
minimal r-mode instability model. However, a probability to
observe even one NS unstable is almost vanishing within the
minimal r-mode instability model (see section
\ref{Sec_minmod}). Thus, observations of a large number of
NSs, which appear to be unstable, can be treated as
evidence of additional (apart from the shear viscosity)
damping at the temperatures of these objects, $T^\infty\sim
3\times10^7-10^8$~K, or r-mode saturation at very small
saturation amplitude (e.g.,
\citealt{hah11,hdh12,gck14b,ms13}). Note that this problem
can hardly be resolved by enhancing the shear viscosity
(e.g., to stabilize 4U 1608--522 one should enhance it by a
factor of 1000,
{and we are unaware of any physically motivated mechanism
predicting such a strong enhancement})
 or by fitting the mass-radius relation
[for example,
{
4U 1608--522 should be stable according to equations
(\ref{tau_GW})--(\ref{tau_S}) if radius $R \lesssim 5$~km.
However,
these equations are written for a
baryonic
equation of state,
and thus they are
inapplicable for such a small radius, because such equations of state predict larger radii (see,
e.g., \citealt*{Oertel_etal16,Zdunik_etal16}); observations
also favour larger NS radii (e.g.,
\citealt{of16_Mass_and_Radius})].}

\subsection{Timing of MSPs}
\label{Sec_timing}

Timing of MSPs allows one to measure the spin frequency
(see ticks on the left side of $\nu$ axis in Fig.\
\ref{Fig_MinMod}; only pulsars with  $\nu>33$~Hz are shown
in the figure) and the spin-down rate. These data were
taken from ATNF Pulsar
Catalogue (\citealt{atnf_paper}%
{, catalogue version 1.55 was applied}).%
\footnote{\url{http://www.atnf.csiro.au/research/pulsar/psrcat/expert.html}}

Within the minimal r-mode instability model MSP spin
frequencies should be limited by the r-mode instability.
Thus, the origin of the high frequency MSPs (especially those
with $\nu\gtrsim 600$~Hz) is unclear in this model, because
they should have very low equilibrium temperatures
{($T^\infty\lesssim 10^7$\,K)}
to be stable during the spin up
in the LMXB stage
{(see Fig.\ \ref{Fig_MinMod}). As shown in sections
\ref{Sec_broad} and \ref{Sec_gen}, such low temperatures
are incompatible with existing accretion and cooling
models.}
Thus, {formation} of MSPs requires additional damping or
use of a tiny saturation amplitude model (see section
\ref{Sec_tiny} and, e.g., \citealt{as14_msp}).

{Another}
constraint on the r-mode instability comes from the
internal heating mechanisms -- superfluid vortex creep
(\citealt{Alpar_etal84}), rotochemical heating
(\citealt{Reisenegger95}), and rotation-induced deep
crustal heating (\citealt{gkr15}) -- which can be efficient
enough to push PSR B1937+21 (J1939+2134 in J2000 system)
into the r-mode instability window (see Fig.\
\ref{Fig_MinMod}). To estimate the internal heating power
$W^\mathrm{i}$  for MSPs we rely on the rotation-induced
deep crustal heating mechanism, because it is not affected
by the uncertainties in the parameters of baryon
superfluidity. Physics of rotation-induced deep crustal
heating is rather simple. MSP spin-down (e.g., by
magneto-dipole losses) results in a gradual compression of
stellar matter. Within the recycling scenario paradigm, the
crust of MSPs should consist of accreted matter, thus the
compression should lead to the same nuclear transformations
in the crust as in the case of accreting NSs (i.e., to deep
crustal heating). For example, for PSR B1937+21 which has
high frequency ($\nu\approx 642$~Hz) and relatively fast
spin-down rate ($\dot \nu\approx 4.3\times
10^{-14}$~s$^{-2}$), the rotation-induced deep crustal
heating {(which is generally proportional to $W^i\propto
\nu \dot{\nu}$)} gives
 $W^\mathrm{i}\sim (10^{31}- 2\times 10^{32})$~erg\,s$^{-1}$,
depending on the equation of state
and pulsar mass.
 To balance
this heating power at the lowest possible temperature (in
order to get it stable), a powerful cooling process should
be active in the core of this source. To get an impression
of the possible realistic cooling luminosity, we adopt here
a simple phenomenological model, which assumes that the
unsuppressed hyperon direct URCA process (see, e.g.,
\citealt{ykgh01}) is active in the central region of the
core with the radius $R_\mathrm{D}$, where hyperons are
present (see \citealt*{kgc16} for a detailed motivation).
The corresponding luminosity is%
%
\footnote{Observations of SAX J1808.4$-$3658 ($\nu=401$~Hz)
can indicate that even more violent cooling is possible.
The observations put an upper limit on the internal
temperature $T^\infty<1.1\times 10^7$~K (e.g.,
\citealt{gck14b}), but the deep crustal heating produces
$\sim 5\times 10^{32}$~erg\,s$^{-1}$ for the average
accretion rate $\dot M\sim 9\times 10^{-12}\ M_\odot$/yr,
reported by \cite{heinke_et_al_09}.}
%
%
\begin{eqnarray}
    L^\mathrm{URCA}_\mathrm{cool}&\approx&  10^{33}\,
    \mathrm{erg\,s}^{-1}
    \left(\frac{T^\infty  }{2\times10^7\, \mathrm{K}}\right)^{6}
    \left(\frac{R_\mathrm{D}}{3\,\mathrm{km}}\right)^{3}.
    \label{L_DURCA}
\end{eqnarray}
For $R_\mathrm{D}=3$~km the equilibrium temperature of PSR
B1937+21, at which $W^\mathrm{i} \approx L_{\rm cool}^{\rm
DURCA}$, can be estimated as
{ $T^{\infty}\sim (0.9- 1.5)\times 10^7$~K.} In the absence
of direct URCA, the equilibrium temperature becomes larger,
so that the source is well inside the instability window
(see Fig. \ref{Fig_MinMod}, where it is shown by the star
symbol). Note that some other internal heating mechanisms
(see, e.g., \citealt{gr10} for a review) can be competitive
with the rotation-induced deep crustal heating, further
increasing the temperature of PSR B1937$+$21 and making it
even more unstable. However, as explained in section
\ref{Sec_minmod}, r-modes rapidly heat up an unstable star
(on a timescale of a century) and spin it down. Thus,
observations of PSR B1937$+$21 contradict the minimal
r-mode instability model: To explain this source one should
stabilize it by introducing an additional damping
mechanism,
{which prevents
r-modes from being unstable.}

{Summarizing, current state of high-frequency MSPs can be
explained within the minimal r-mode instability model only
if they are cold enough to be stable. However, this
explanation has two problems: (i) MSPs cannot be produced
by accretion at such low temperatures; (ii)
PSR B1937$+$21 cannot be in thermal equilibrium inside the stability region
(cooling cannot compensate internal heating).
As discussed below, both
problems can be resolved within the tiny r-mode saturation
amplitude scenario
(but then the required saturation amplitudes are
unfeasible, at least for hadronic equations of state,
see section \ref{Sec_tiny})
or by additional dissipation of r-modes at $T^\infty\sim 2\times 10^7$~K
(sections \ref{Sec_bulk} -- \ref{Sec_gen}).}

\subsection{X-ray and \textit{UV} observations of MSPs}
\label{Sec_Xray_MSP}

The NS temperature is a crucial parameter for the r-mode
instability and hence any observational constraint on the
stellar temperature can be very important. A large number
of MSPs has been observed in X-rays, however, the thermal
emission from them has been either not detected or detected
only from pulsars' hot spots
(e.g., \citealt{Zavlin07,Bogdanov_etal11}). Thermal emission
from the \underline{whole} surface is typically not
required to fit the X-ray spectrum. A notable
{exceptions are  PSR J0437$-$4715 ($\nu\approx 173.7$~Hz)
and PSR J2124$-$3358 ($\nu\approx 202.8$~Hz)} for which
combined analysis of \textit{UV} and X-ray spectra
allows \cite{Durant_etal12_0437} and \cite{Rangelov_etal17} to estimate the surface
temperature of these sources:
$T^\infty_\mathrm{eff}\sim (1-3)\times 10^5$~K
 and
$T^\infty_\mathrm{eff}\sim (0.5-2.1)\times 10^5$~K,
 respectively. The
corresponding internal temperature is $T^\infty\lesssim
10^7$~K and thus
{these pulsars are} r-mode stable
{(even within the minimal r-mode instability model).}

\begin{figure}
        \includegraphics[width=\columnwidth]{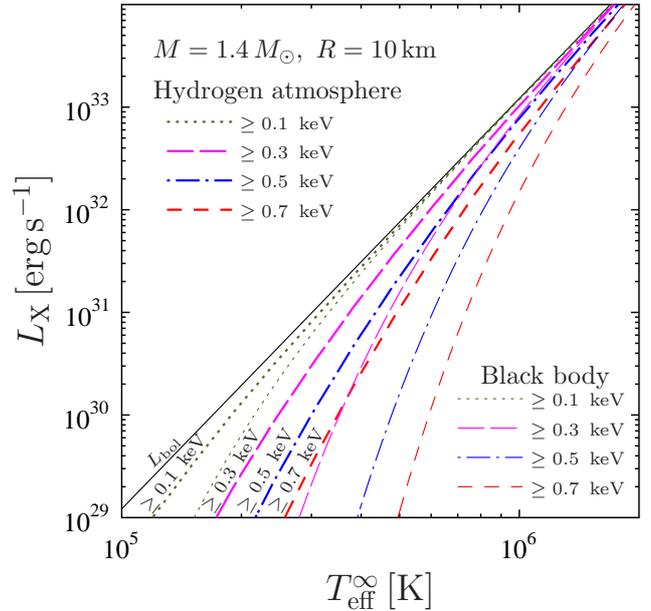}
   \caption{{Thermal X-ray luminosity (as seen by a distant observer) from the whole surface of a canonical NS
    as a function of the effective surface temperature.
    Bolometric luminosity is shown by solid line, and
    other lines correspond to energy ranges
    $\ge 0.1,\ 0.3,\ 0.5$, and $0.7$\,keV.
    Thick lines are for the hydrogen atmosphere model
    and thin lines -- for the black-body model. }}
    \label{Fig_Lum}
\end{figure}

\cite{Schwenzer_etal_Xray} have recently suggested a strong
upper limit, $T_\mathrm{bb}<1.7\times 10^5$~K, on the
surface temperature of PSR J1231-1411 ($\nu\approx
271.5$\,Hz) within the blackbody model. As the authors
emphasize, they do not account for gravitational redshift
and ignore the effects related to stellar atmosphere, which
can, in our opinion, significantly affect the actual value
of the upper limit (see, e.g., \citealt*{zps96}, %
{Fig.\ \ref{Fig_Lum} and discussion below%
}).
\cite{Schwenzer_etal_Xray} have also proposed upper limits
on the surface temperature of several other MSPs based
either on estimates of their total
{(or, in several cases, thermal)} luminosity (available in
the literature) or on the results of \cite{pb15}, who
estimated maximum possible $T_\mathrm{bb}$ for a number of
sources.

Here we suggest a general upper limit on the surface
temperatures of MSPs. For several pulsars (e.g., for PSR
1937+21) our upper limit is stronger than that suggested by
\cite{Schwenzer_etal_Xray}.

{Indeed, } the MSP surface temperatures are strongly
constrained by existing observations: If MSPs were as hot
as transiently accreting NSs, whose thermal emission is
observed in X-rays even in the quiescent state (see section
\ref{Sec_Xray} and Fig.\ \ref{Fig_MinMod}), the thermal
emission from the whole MSP surface would be detected in
X-rays as well. But it is not the case. Thus, leaving a
detailed analysis of X-ray observations for each MSP beyond
the scope of this paper, we assume that MSPs should
(generally) be colder than the transiently accreting NSs
with the measured surface temperature. To get an idea of
the possible upper limit on the MSP surface temperature we
rely on observations of MXB 1659$-$298 (assumed distance to
the source $D=10$~kpc and X-ray absorption column density
$N_\mathrm{H}=(2.0\pm 0.2)\times 10^{21}$~cm$^{-2}$),
 IGR 00291+5934 ($D=4$~kpc and $N_\mathrm{H}=2.8\times 10^{21}$~cm$^{-2}$),
and HETE J1900.1--2455 ($D\approx 4.7$~kpc and
$N_\mathrm{H}\approx 2.2\times 10^{21}$~cm$^{-2}$),
reported by \cite{cackett_et_al_08},
\cite{heinke_et_al_09},
 and \cite{Degenaar_etal16_J1900}.
For all of them the effective surface temperature was
estimated as $T_\mathrm{eff}^\infty\approx 6.3\times
10^5$~K. Note that, for these objects both absorption
column density and distances are relatively large in
comparison to
typical
fast MSPs (with $\nu>250$~Hz).%
%
\footnote{ \cite*{hnk13} propose the following estimate for
the absorption column density $N_\mathrm{H}$:
$N_\mathrm{H}=(0.30^{+0.13}_{-0.09})\times
10^{20}\,\mathrm{cm} \,(\mathrm{DM}/\mathrm{pc\,cm}^{-3})$,
where $\mathrm{DM}$ is the dispersion measure. Thus, MSPs
with $\mathrm{DM}<70\,\mathrm{pc\,cm}^{-3}$ should have
$N_\mathrm{H}\lesssim 2.1\times 10^{21}$~cm$^{-2}$, a
requirement, which is satisfied for more than
{$67\%$} of currently known fast MSPs ($\nu>250$~Hz),
according to ATNF
pulsar catalogue (\citealt{atnf_paper}%
{, catalogue version 1.55 was applied}%
). This catalogue
also gives an information on the pulsar distances, and it
turns out that 70\% of known fast MSPs ($\nu>250$~Hz) are
located closer than HETE J1900.1--2455, but this statistic
is not independent because the pulsar distances are
estimated using the
\cite{ymw17}
DM-based distance model as
default. }
%
Thus, it is unlikely that the lack of successful observations
of thermal emission from the whole surface of fast MSPs can
be explained by strong absorption or large distances to
these sources: Rather, it is more probable that they are
`cold' (colder than MXB 1659$-$298,
 IGR 00291+5934,
and HETE J1900.1--2455). Therefore, we choose
\begin{equation}
 T_\mathrm{eff}^\infty< \teffmax= 6\times 10^5 \, \mathrm{ K}
 \label{Teffmax}
\end{equation}
as a fiducial upper limit for the effective surface
temperature of MSPs. This temperature corresponds to the
bolometric thermal luminosity $L_\mathrm{bol}\lesssim
1.6\times 10^{32}$~erg\,s$^{-1}$ [see equation (\ref{L_ph})
below].
For photon energy ranges, suitable for X-ray observations
($>0.1,\
>0.3$, and
$>0.5$\,keV), $T_\mathrm{eff}^\infty=6\times 10^5 \,
\mathrm{ K}$ corresponds to lower luminosity:
$L^\mathrm{>0.1}_\mathrm{X}=1.5\times 10^{32},$\
$L^\mathrm{>0.3}_\mathrm{X}=1.0\times10^{32}$, and
$L^\mathrm{>0.5}_\mathrm{X}=0.6\times10^{32}$\,erg\,s$^{-1}$,
respectively (see also Fig.\ \ref{Fig_Lum} which represents
X-ray luminosity as a function of $T_{\rm eff}^\infty$). Here
and below the upper index (e.g., $^{>0.3}$) denotes the
photon energy range (in keV). The quoted numbers are
insensitive to the upper bound of the photon energy range
(if it exceeds 1\,keV) and are calculated for a canonical
NS with hydrogen atmosphere ({\tt XSPEC} models
{\tt nsa} by \citealt{zps96} and {\tt nsatmos} by
\citealt{hrng06} were applied and give the same result).
The black-body model has the same bolometric luminosity
(by definition of the effective temperature), but the
luminosity in the X-ray part of the spectrum is much lower
(compare thick and thin lines in Fig.\ \ref{Fig_Lum} and
see, e.g., \citealt{zps96} for details).

Polar cap and magnetospheric emission can only increase MSP
luminosity, thus total unabsorbed X-ray luminosity,
extracted from observations, can be applied to constrain
$T_{\rm eff}^\infty$: thermal contribution can not exceed
the total luminosity (in any energy range). For example, if
total unabsorbed X-ray emission from some MSP is
$L^\mathrm{>0.3}_\mathrm{X}<1.0\times10^{32}$\,erg\,s$^{-1}$,
its effective temperature should be $T_\mathrm{eff}^\infty<
6\times 10^5 \, \mathrm{ K}$, confirming our constraint
(\ref{Teffmax}) for it. Note that, the black body emission
model gives less strict constraints (see thin lines in
Fig.\ \ref{Fig_Lum}; for example
$L^\mathrm{>0.3}_\mathrm{X}<1.0\times10^{32}$\,erg\,s$^{-1}$
translates into $T_\mathrm{eff}^\infty< 6.7\times 10^5 \,
\mathrm{ K}$).

{Let us confront this criterion with observations.
$L_\mathrm{X}$ was estimated for a rather large number of
MSPs and, typically, $L^{(0.3-8)}_\mathrm{X}\sim 4\times
10^{30}-10^{31}$\,erg\,s$^{-1}$
(\citealt{Bogdanov_etal11_M28}), which gives
$T^\infty_\mathrm{eff}\lesssim (3.2-3.5)\times10^5$\,K
(assuming a canonical NS with hydrogen atmosphere), i.e.\
smaller than our fiducial upper limit (\ref{Teffmax}). MSPs
in globular cluster 47 Tuc provide yet another example
suggesting that the observations put even stronger
constraint on $T_{\rm eff}^\infty$ than (\ref{Teffmax}).
This globular cluster is particularly good for X-ray
studies, because it is close ($D\approx 4.5$~kpc) and
weakly absorbed (cluster $N_H\approx 1.3\times
10^{20}$~cm$^{-2}$, e.g.,
\citealt{heinke_etal05_Chandra_47Tuc}).
\cite{bogdanov06_19MSP_47Tuc} found X-ray counterparts for
all 19 MSPs with precisely known positions in this globular
cluster, and X-ray luminosity for all of them [including 47
Tuc J ($\nu\approx 476$~Hz), Y ($\nu\approx 455$~Hz), and W
($\nu\approx 425$~Hz),
which were the fastest MSPs in this globular cluster known at that moment]%
\footnote{Recently, \cite{pan_etal16} discovered a new
pulsar in this cluster, PSR J0024-7204aa, which has a higher
frequency, $\nu\approx 541$~Hz.}
agrees with the constraint $L^{(0.3-8)}_\mathrm{X}\lesssim
3\times 10^{31}$\,erg\,s$^{-1}$,
corresponding to $T_\mathrm{eff}^\infty\lesssim 4.7\times 10^5$~K.
For pulsar 47 Tuc Y the flux, listed in table 3 of
\cite{bogdanov06_19MSP_47Tuc},
corresponds to
$L^{(0.3-8)}_\mathrm{X}= 3\times 10^{30}$\,erg\,s$^{-1}$,
setting the following upper limit on
the surface temperate of this pulsar,
$T_\mathrm{eff}^\infty\lesssim 3\times 10^5$\,K.
Finally, X-ray
luminosity, $L^{0.3-8}_\mathrm{X}=10^{31}$\,erg\,s$^{-1}$,
reported
for PSR J1810+1744 ($\nu\approx 602$~Hz)
by \cite{Gentile_etal14_XRayBlackWidows},
implies
$T_\mathrm{eff}^\infty\lesssim 3.8\times
10^5$\,K. }

%
{ Some MSPs have stronger X-ray emission. For example, for
PSR B1957+20 (also known as J1959+2048, $\nu\approx
622$\,Hz) the unabsorbed X-ray flux in $(0.15-10)$\,keV
range was estimated by \cite*{apg15_BW} as $6.5\times
10^{-14}$\,erg\,cm$^{-2}$\,s$^{-1}$, which corresponds to
$L^{0.15-10}_\mathrm{X}=5^{+5}_{-3}\times
10^{31}$\,erg\,s$^{-1}$, for $D=2.5\pm 1$\,kpc
(\citealt{Huang_etal12_B1957}, inferred from the dispersion
measure model of \citealt{cl02_distModel}).
Even the upper bound of this
luminosity, $L^{0.15-10}_\mathrm{X}=10^{32}$\,erg\,s$^{-1}$,
constrains the temperature of PSR B1957+20 at
$T_\mathrm{eff}^\infty\lesssim 5.6\times 10^5$\,K, in
agreement with (\ref{Teffmax}).}

{
 The X-ray luminosity of MSPs with high spin-down
power
can be
even larger.
However, they have power-law X-ray spectrum
and a high fraction of pulsed emission
(see, e.g., section 4.2 in \citealt{Ng_etal14_B1937}).
In this paper we are
interested in the internal temperatures, which reveal
itself in the thermal emission from the whole surface,
contributing to the unpulsed component of X-ray emission.
Thus, the analysis of the phase-averaged spectra can lead to
overestimation of $T^\infty_\mathrm{eff}$.
For example,
for PSR B1937+21,
which plays an especially important role in
this paper,
$L^{0.5-7}_\mathrm{X}\approx 6.8\times 10^{32}$~erg\,s$^{-1}$,
but, as stated by \cite{Ng_etal14_B1937}, it
shows
`purely non-thermal spectrum with a hard photon index of
$0.9\pm0.1$, and is nearly 100\% pulsed'.
Thus the thermal
contribution to the unpulsed luminosity should be much smaller.
Unfortunately, numerical upper bound is not
reported by \cite{Ng_etal14_B1937}, but violation of the
constraint (\ref{Teffmax})
would lead
to X-ray luminosity
$L^{(0.5-7)}_\mathrm{X}>0.6\times10^{32}$\,erg\,s$^{-1}$,
corresponding to the thermal contribution
of
the unpulsed
fraction of X-ray emission $\gtrsim 9\%$,
which exceeds an error in the determination of the unpulsed fraction of X-ray
emission for B1821-24 and J0218+4232 ($4\%$ and $6\%$,
respectively; \citealt{Ng_etal14_B1937}).
Thus, we suppose
that such a
strong fraction of the unpulsed emission is
excluded, but this conclusion should be checked by accurate
analysis of the X-ray data. }
 Note that
\cite{Schwenzer_etal_Xray} have
used total X-ray
luminosity ($L_\mathrm{X}\approx 6.88\times
10^{32}$~erg\,s$^{-1}$, quoted in their table 1) to obtain
an upper limit on the surface temperature of PSR B1937+21.
It can be a reason why their result
($T_\mathrm{eff}^\infty\sim 10^6$\,K, according to their
figure 1) exceeds (\ref{Teffmax}). High upper bounds for
the surface temperature of some MSPs reported by
\cite{pb15} (also quoted in table 2 in
\citealt{Schwenzer_etal_Xray}) correspond to pulsars
undetected in X-rays; thus they can not be considered as
evidence of violation of (\ref{Teffmax}).

{Summarizing, for faint X-ray pulsars
[$L^{(0.3-8)}_\mathrm{X}\lesssim 10^{31}$\,erg\,s$^{-1}$,
e.g., 47 Tuc Y $\nu\approx 455$~Hz and PSR J1810+1744
$\nu\approx 602$~Hz],
total X-ray luminosity allows us to
limit surface temperature at
$T_\mathrm{eff}^\infty\lesssim(3-4)\times 10^5 \, \mathrm{ K}$.
For more luminous pulsars [$L^{(0.3-8)}_\mathrm{X}\lesssim
10^{32}$\,erg\,s$^{-1}$, e.g., PSR B1957+20 ($\nu\approx
622$\,Hz)]
$T_\mathrm{eff}^\infty$ is constrained by our fiducial limit (\ref{Teffmax}).
The most luminous pulsars
[$L^{(0.3-8)}_\mathrm{X}> 10^{32}$\,erg\,s$^{-1}$, e.g.,\
PSR B1937+21 ($\nu\approx 642$~Hz)] typically have strong
pulsed fraction,
while
$T_{\rm eff}^\infty$
should be
estimated using the
\underline{unpulsed} fraction of
X-ray
luminosity.
We argue, that nearly $\sim100\%$
pulsed emission from PSR B1937+21,
reported by \cite{Ng_etal14_B1937},
supports the constraint (\ref{Teffmax}).
An accurate and detailed
analysis of X-ray and \textit{UV} observational data can lead to
even
tighter upper bounds
on the temperatures of
MSPs.
These bounds will allow
one to apply the analysis of sections
\ref{Sec_tiny}--\ref{Sec_gen} in order to extract more information
about the properties of these sources.
However, in this
paper we shall adopt the conservative general upper limit
(\ref{Teffmax}) and demonstrate that even it constrains the
r-mode instability window.}

{ For NSs with accreted envelope composition (a reasonable
option, since they were recycled by accretion), equation
(\ref{Teffmax}) constraints  the redshifted internal
stellar temperature as
\begin{equation}
T^\infty\lesssim\tmax\approx 2\times 10^7 \, {\rm K}
\label{Tmax}
\end{equation}
(see the hatched region in Fig.\ \ref{Fig_MinMod}). Purely iron
thermally insulating envelope corresponds to a less strict
constraint
\begin{equation}
T^\infty\lesssim T^\infty_\mathrm{max\,Fe}\approx 4.7\times
10^7 \, {\rm K} \label{Tmax_Fe}.
\end{equation}
However, for low temperatures discussed here, an NS model
with even very thin layer of accreted (light) elements strongly
affects $T^\infty_\mathrm{eff}/T^\infty$ ratio (see, e.g.,
Fig.\ 8 in \citealt{pcy97}).
Indeed, almost vanishing
layer of light elements on top of the iron envelope with the mass
$\Delta M=4\times10^{-14} M_\odot$ reduces the estimate
(\ref{Tmax_Fe})  to $T^\infty\lesssim 3\times 10^7 \, {\rm K}$,
while
$\Delta M\gtrsim 10^{-11}M_\odot$ makes it
almost
indistinguishable from the fully accreted result
(\ref{Tmax}).%
\footnote{{For $T_\mathrm{eff}^\infty\lesssim 3\times 10^5$\,K,
which is a reasonable constraint for pulsar 47 Tuc Y, the
internal temperature should be $T^\infty\lesssim 1.3\times
10^7 \, {\rm K}$ even for iron thermally insulating
envelope and $T^\infty\lesssim 7\times 10^6 \, {\rm K}$ for
the layer of light elements $\Delta M>10^{-14} M_\odot$.}}
Thus, we apply Eq.\ (\ref{Tmax}) in what follows.}

{ Strictly speaking, equation (\ref{Tmax}) does not reveal
any new inconsistency in the minimal r-mode instability
model: as discussed in section \ref{Sec_timing}, the
current state of MSPs can be explained within the minimal
model, assuming that they are stable, and thus even colder
(see Fig.\ \ref{Fig_MinMod}). However, MSP formation and
intensive internal heating of PSR B1937$+$21 require a
modification of the minimal model.
As we demonstrate below,
the internal temperature upper bound (\ref{Tmax}) provides an important
constraint on the r-mode instability models which
are able
to solve the MSP formation problem (see sections \ref{Sec_tiny}
-- \ref{Sec_gen}).}

\section{Tiny r-mode amplitude scenario}
\label{Sec_tiny}

Within the tiny r-mode amplitude scenario, NSs can be
unstable, but saturation amplitudes are chosen small enough
to agree with observations. \cite{ms13} confronted this
model with the observations of NSs in LMXBs and concluded
that the r-mode amplitudes should range from $10^{-8}$ to
$1.5\times 10^{-6}$.

\cite{as15_oscMSP}
apply MSP timing
to constrain r-mode amplitudes in MSPs.%
%
\footnote{\cite{as15_oscMSP} have not taken into account
the correction due to Shklovskii (proper motion) effect
(\citealt{shklovskii70}), which can significantly affect
$\dot{\nu}$ for some MSPs with the slowest spin-down rate
(e.g., \citealt{Guillemot_etal16_Shklovskii}).
\label{Shklovskii}}
%
Assuming power-law dependence of relevant parameters on
$\nu$, $T^\infty$, $M$, and $R$, they suggest `universal
r-mode spin-down limit'\ on gravitational wave signal from
MSPs. However, at least for neutrino luminosity this
assumption seems to be oversimplified: Simulations of
stellar cooling and quiescent temperatures of accreting NSs
require different cooling mechanisms to be active in NSs
with different masses leading to very strong
(non-power-law) dependence of luminosity on the mass and
temperature
 [see, e.g.,
\cite{yp04,plps04, lh07,by15a,by15b}]. Thus, universal
r-mode spin-down limit suggested by \cite{as15_oscMSP}
should be treated with a bit of caution.
Furthermore, as we demonstrate below, upper boundary
(\ref{Teffmax}) for the surface temperature of MSPs can
constrain the r-mode amplitudes much stronger. This is in
line with \cite{Schwenzer_etal_Xray} who have recently come
to the similar conclusion.

For saturated r-mode, the heating power can be estimated
from equation (\ref{HeatRate}). In thermal equilibrium the
total heating (sum of all heating mechanisms: r-mode,
rotation-induced deep crustal heating, rotochemical
heating, etc.) should be compensated by cooling. If an MSP
is not massive enough to open direct URCA process in its
core, the neutrino emission
{becomes} negligible [due to temperature constraint
(\ref{Tmax})], so that the total cooling luminosity can be
estimated as the thermal emission from the MSP surface
\begin{eqnarray}
    L^\mathrm{ph}_\mathrm{cool}&=&\frac{4\pi R^2 \sigma
    (T^\infty_\mathrm{eff})^4}{1-2GM/(c^2R)}
     \label{L_ph}
  \\&\approx&
    1.6\times 10^{32}\,   \frac{\mathrm{erg}}{\mathrm{s}}\,R_{10}^{2}
    \left(\frac{T_\mathrm{eff}^\infty}{6\times10^5\ \mathrm{K}}\right)^{4} ,
    \nonumber
\end{eqnarray}
where $\sigma$ is the Stefan-Boltzmann constant. Here and
below the red-shift factor, $1+z=1/\sqrt{1-2GM/(c^2R)}$, is
assumed to be $1.3$. This value corresponds to the
canonical NS. The upper limit on the r-mode amplitude can
be computed assuming that r-modes are the only source of
heat
(see equation \ref{HeatRate}).%
\footnote{Coefficient in equation (6) by
\cite{Schwenzer_etal_Xray} differs from ours because it
corresponds to $M=M_\odot$, $R=10$~km, and
$L^\mathrm{ph}_\mathrm{cool}=4\pi\, \sigma \, R^2
T_\mathrm{eff}^4$, where
$T_\mathrm{eff}=(1+z)\,T^\infty_\mathrm{eff}$ is
the unredshifted (local) effective temperature.}
\begin{equation}
   \alpha<\alpha^\mathrm{ph}=3\times 10^{-8}
   \, R_{10}^{-2}\,
    M_{1.4}^{-1}\,
    \nu_{500}^{-4}\,
    \left(\frac{T_\mathrm{eff}^\infty}{6\times 10^5\ \mathrm{K}}\right)^{2}.
    \label{a_ph}
\end{equation}
For massive NSs the direct URCA processes can operate in
the central region of the core.
{ Plugging the relation between the surface and internal
temperatures for fully accreted envelope (\citealt{pcy97})
into expression (\ref{L_DURCA}) for the neutrino
luminosity, we obtain: }
\begin{equation}
    L^\mathrm{URCA}_\mathrm{cool}
    \approx
    9\times10^{32}    \frac{\mathrm{erg}}{\mathrm{s}}
    \left(\frac{T_\mathrm{eff}^\infty}{6\times 10^5\,\mathrm{K}}\right)^{9.92}
    \,\left(\frac{R_\mathrm{D}}{3\,\mathrm{km}}\right)^{3}
    \left(\frac{g_\mathrm{c}}{g}\right)^{2.48},
     \label{L_DURCA_Teff}
\end{equation}
where
\begin{equation}
    g=\frac{GM}{R^2\sqrt{1-2GM/(c^2R)}}
    \label{g}
\end{equation}
is the surface gravity (for the canonical NS
$g=g_\mathrm{c}=2.43\times10^{14}\,\mathrm{cm/s}^2$). The
luminosity (\ref{L_DURCA_Teff}) results in a generally
weaker constraint on the r-mode amplitude,
\begin{eqnarray}
\alpha<\alpha^\mathrm{URCA}&=&8\times 10^{-8}\,
    \nu_{500}^{-4}\,
        R_{10}^{-0.52}
        \,M_{1.4}^{-2.24}
\nonumber
    \\
    &\times&
     \left(\frac{R_\mathrm{D}}{3\,\mathrm{km}}\right)^{1.5}
    \,
    \left(\frac{T_\mathrm{eff}^\infty}{6\times 10^5\
    \mathrm{K}}\right)^{4.96}.
       \label{a_DURCA}
\end{eqnarray}
If one allows for both cooling processes (photon and direct
URCA), one finds the following  upper limit on the r-mode
amplitude,
\begin{equation}
\alpha<a^\mathrm{gen}=\sqrt{{(\alpha^\mathrm{URCA})}^2+{(\alpha^\mathrm{ph})}^2}.
\label{a_gen}
\end{equation}
Note that the constraint (\ref{a_gen}) is tighter than the
upper bound found by \cite{as15_oscMSP} from the analysis
of MSP spin-down rate. Thus, r-mode can be responsible only
for a small fraction of the total spin down rate  $\dot
\Omega^\mathrm{tot}= 2\pi\dot \nu^\mathrm{tot}$ (see
\citealt{Schwenzer_etal_Xray} for a detailed discussion).

{It is worth noting that,} for
$T^\infty_\mathrm{eff}\lesssim 3\times 10^5$~K {[an upper
limit for the surface temperature of PSR J0024$-$7204Y
(=47Tuc Y)]},
the photon emission from the stellar surface dominates.
Thus, {in that case an} upper bound on the r-mode amplitude
{follows} {\it directly from observations} [by using Eq.
(\ref{a_ph})], without referencing to rather uncertain (but
 negligible at $T^\infty_\mathrm{eff}\lesssim 3\times
10^5$~K) neutrino emission from the star.

{Note also that the saturation amplitude generally
increases with the temperature decrease (e.g.,
\citealt*{bw13}
%
\footnote{See, however, footnote 9 in \cite{gck14b} for a critique of the low-saturation amplitude
model of \cite{bw13}.}%
%
).
Thus, an upper limit for the
low-temperature MSPs is, in fact, a stronger constraint
than numerically the same upper limit for hot NSs in LMXBs.
In any case, the upper limit (\ref{a_gen}) on the r-mode
amplitude in MSPs is too strong to be explained by
previously suggested mechanisms of nonlinear saturation of
r-modes ({namelly, saturation by nonlinear mode couplings discussed by \citealt{Arras_etal03,btw07,bw13} and by vortex/flux tube interaction suggested by \citealt{hga14}).
The only exception is the work by \citealt*{ahs15},
but it assumes a hybrid star with sharp hadron-quark
interface and thus
it
is inapplicable to baryonic equations
of state.
}
Hence, explanation of MSPs within the tiny r-mode model
is unfeasible (at least in the absence of quark core inside NSs).
}

 At the end
of the section it is worth noting that the estimates
obtained above constrain  the {(root-mean-squared)} r-mode
amplitude even if it is limited not by saturation, but by
the shape of the instability window (e.g., within the
resonance uplift scenario; see sections \ref{Sec_reson} and
\ref{Sec_gen}).

\section{Stabilization by the bulk viscosity: Too strong neutrino luminosity}\label{Sec_bulk}

As discussed in section \ref{Sec_observ}, shear viscosity
is not sufficient to stabilize r-modes in the observed NSs.
Can this problem be solved by the bulk viscosity? For
nucleonic equations of state the bulk viscosity is very
small at $T\lesssim 10^8$~K, being negligible in comparison
with the shear viscosity. However, for another core
composition the bulk viscosity can be much larger. For
example, \cite{as14_msp} demonstrates that the model of
ungapped interacting quark matter can successfully
stabilize NSs in LMXBs.
This model predicts that the fastest MSPs (with $\nu\gtrsim
600$~Hz) should have internal temperatures $T\gtrsim
3\times 10^7$~K (see their Fig.\ 1) and thus slightly
contradicts
the internal MSP temperature constraint (\ref{Tmax}).%
%
\footnote{According to Fig.\ 3 of \cite{as14_msp}, the
condition (\ref{Teffmax}) for $T_{\rm eff}^\infty$ is
fulfilled, but this figure corresponds to {purely iron
thermally insulating envelope}, which does not look
realistic in the case of MSPs {(see section
\ref{Sec_Xray_MSP})}. For accreted envelope the ungapped
interacting quark matter stability curve shifts to higher
surface temperatures and violates the constraint
(\ref{Teffmax}).}
%

Probably the more important problem of \cite{as14_msp} is
high neutrino luminosity, $L_\mathrm{cool}^\mathrm{uiq}\sim
4\times 10^{37} (T_8^\infty)^6$~erg\,s$^{-1}$, inherent to
their
model.%
\footnote{To obtain this estimate we apply equation (1)
from \cite{as14_msp}, neglecting the non-Fermi-liquid
effects and assuming that it is written in Plank units.
Following \cite{ams12_small}, we use
$\Lambda_\mathrm{QCD}=1$~GeV,
$\Lambda_\mathrm{EW}=100$~GeV. Note that these constants
should not be confused with the constants denoting
nonperturbative scale of QCD ($340\pm 8$~MeV) and scale of
electroweak symmetry breaking $\approx 246$~GeV (e.g.,
\citealt{pdd14}), which often have the same notation.
}
%
For example, 4U 1608$-$522 is r-mode stable in ungapped
interacting quark model. Thus, the main heat source for
that NS is deep crustal heating. It provides
$W^\mathrm{DCH}\approx 2\times 10^{34}$~erg\,s$^{-1}$ (for
the average accreting rate $\dot M\approx 3.6\times
10^{-10}~\mathrm{M}_\odot/\mathrm{yr}$, see \citealt{hjwt07}),
corresponding to the equilibrium temperature
$T^\infty_\mathrm{eq}\approx 3\times 10^7$~K. However, this
temperature is strongly below the observational estimate,
$T^\infty\approx (0.9-2.5)\times 10^8$~K (see, e.g.,
\citealt{gck14b}). To keep 4U 1608$-$522 at the observed
temperature there should be an additional heat source,
compensating the neutrino luminosity
$L_\mathrm{cool}^\mathrm{uiq}\approx (2\times
10^{37}-10^{40})$~erg\,s$^{-1}$. The nature of such source
is unclear, especially for stable NSs. The interpretation of
temperatures of other hot NSs (with $T^\infty\sim 10^8$~K)
has the same problem.

The enhanced neutrino emission, leading to problems with
explanation of hot NSs in LMXBs, should be common for
models, which stabilize NSs by the bulk viscosity. The
reason is simple: Both neutrino luminosity and bulk
viscosity are associated with the reactions between the
particles in the NS core (see, e.g.,
\citealt*{rb03,gyg05,ams10}). If all these reactions are
accompanied by neutrino production (as it is the case for
nucleonic matter), there should be a direct correspondence
between the bulk viscosity (for small-amplitude
oscillations) and neutrino emissivity (see, for example,
equations 1 and 2 in \citealt{rb03_bulk}, written for the
modified URCA reactions). This means that enhancement of
bulk viscosity unavoidably results in an enhancement of
neutrino luminosity.

For more complicated core compositions (e.g., for
nucleon-hyperon or quark matter) non-leptonic reactions are
possible (e.g., \citealt*{jones01, hly02_adiab}) that do
not involve neutrinos and, being more powerful than
neutrino reactions, produce dominant contribution to the
bulk viscosity (e.g.,
\citealt*{jones01,hly02_bulk,lo02,gk08}, \citealt{ams10}).
Thus, the ratio of the bulk viscosity energy dissipation
rate to luminosity can be enhanced in comparison to the
nucleonic matter.
However, to stabilize the fastest NSs in LMXBs, like 4U
1608-522, bulk viscosity should be enhanced at least by 13
orders of magnitude (cf.\ equations \ref{tau_GW} and
\ref{tau_B}), which seems to be hardly possible without
substantial increase in the neutrino luminosity. At the
same time, if all NSs in LMXBs are r-mode stable, the
quiescent temperatures of the hottest of them can be
explained {\it only} if neutrino luminosity is not too
strong, i.e., it should be on the level of the modified
URCA luminosity for nucleonic cores (see, e.g.,
\citealt{yp04,lh07}).

The above arguments are, of course, not rigorous enough,
but we want to stress that each particular model with
r-mode stabilization by enhanced bulk viscosity should be
checked for consistency with the observed quiescent
temperatures of the hottest NSs in LMXBs (in particular, 4U
1608-522).

\section{Resonance uplift scenario:
low temperature resonances are required} \label{Sec_reson}

\begin{figure*}
       \includegraphics[width=18cm]{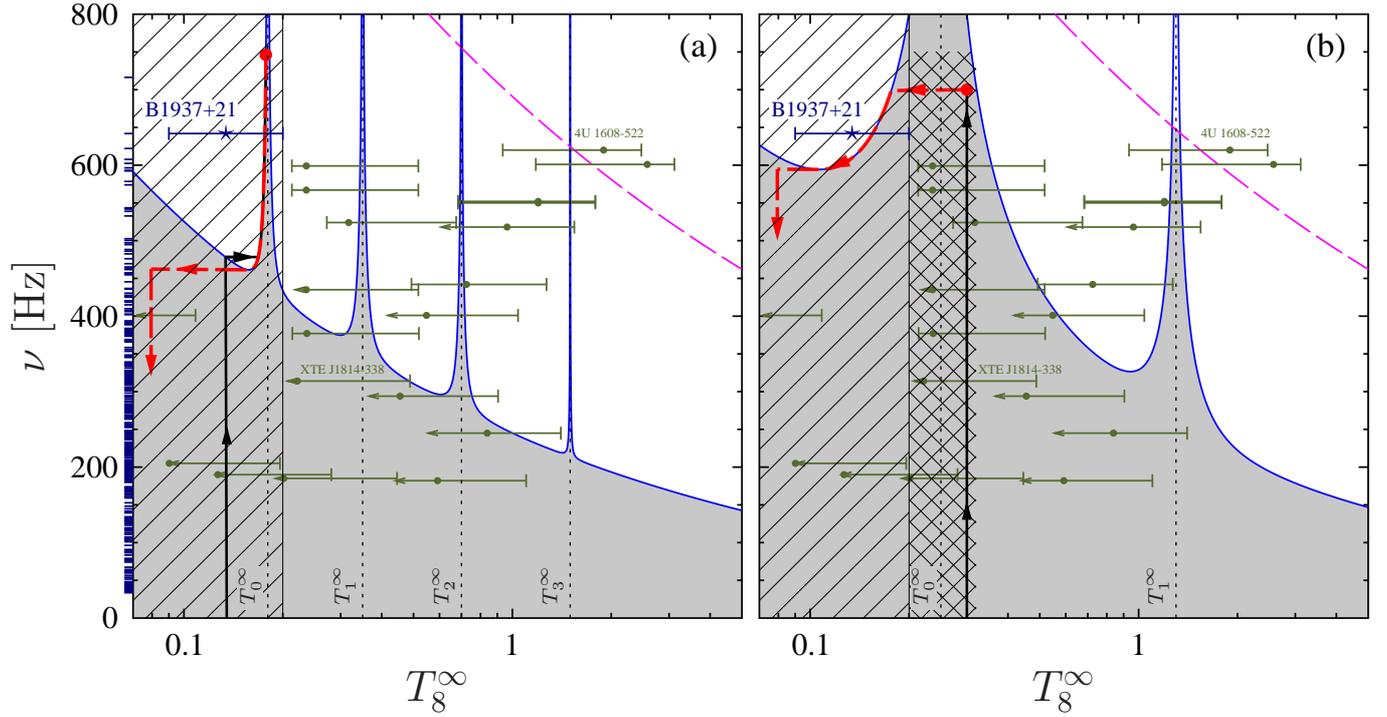}
    \caption{Example of r-mode instability windows in the resonance uplift scenario
    for narrow [panel (a)] and
    broad [panel (b)] stability peaks.
    The dashed line demonstrates
    the instability curve
    for $m=3$ r-mode; all sources higher than this curve are unstable
    (even inside the stability peaks).
    Vertical dotted lines represent resonance temperatures
    $T^\infty_i$ for each stability peak.
    Typical evolution path for an accreting NS in LMXB
    is shown by a thick solid line.
    Evolution path for a NS at the MSP stage is shown by thick dashed line.
    Filled circle marks the NS frequency and temperature immediately after the Roche-Lobe
    decoupling.
    Minimal stability region
    required for the formation of MSPs via recycling scenario is cross-hatched
    (see section \ref{Sec_gen} for details).
    Other notations are the same as in Fig.\ \ref{Fig_MinMod}.
    }
    \label{Fig_UpLift}
\end{figure*}

Resonance uplift scenario was suggested by
\cite{gck14a,gck14b} to explain observations of NSs in
LMXBs. It is based on the proper account of finite
temperature effects on the dynamics of superfluid liquid in
the core and predicts an enhanced NS stability associated
with the resonance interaction of oscillation modes in the
vicinity of certain (resonance) temperatures (see Fig.\
\ref{Fig_UpLift}, where the instability window is splitted
by stability peaks). As discussed by \cite{gck14a,gck14b},
this picture agrees with detailed calculations of
frequencies and damping times for: (i) r-modes and inertial
modes of superfluid NSs at zero temperature
(\citealt{ly03,yl03a}); and
(ii) f- and p-modes in nonrotating superfluid NSs at finite
temperatures (\citealt{ga06, kg11,cg11,gkcg13,gkgc14}).
Furthermore, \cite{kg16} confirm the main features of the
resonance uplift scenario by calculating the r-mode
spectrum and damping times for a finite-temperature
superfluid NS within a simplified model with vanishing
entrainment. However, the form of the stability peaks
depends on many parameters (e.g., on the critical
temperature profiles, equation of state etc.) and remains
rather uncertain. Here we shall try to constrain it by
making use of the observations discussed in section
\ref{Sec_observ}.

As argued in \cite{gck14b}, the fastest MSPs should be
attached to the left slopes of some of the existing
stability peaks. Note that those peaks (more accurately,
their left slopes) should appear at $T^\infty\lesssim
\tmax$ in order to match the observational upper limit on
MSP temperature (\ref{Tmax}). Here, within the
phenomenological approach, we discuss two models: (a) a
model with {\it narrow} low-temperature stability peaks,
for which the cooling luminosity $L_\mathrm{cool}$ does not
change significantly inside the peak [see panel (a) in
Fig.\ \ref{Fig_UpLift} and section \ref{Sec_narrow}] and (b) a
model with {\it broad} low-temperature stability peaks [see
panel (b) and section \ref{Sec_broad}].

Formally, both models can explain the current state of the
observed rapidly rotating NSs: each NS in an LMXB is either
attached to the left slope of some stability peak (e.g., 4U
1608-522) or stable with respect to r-modes (e.g., XTE
J1814-338), as it should be in the resonance uplift
scenario (e.g., \citealt{gck14b}). High-frequency MSPs
[with $\nu\gtrsim 450$~Hz for model (a) and $\nu\gtrsim
600$~Hz for model (b)] can be explained as moving along the
left slope of the low-temperature stability peak, other
MSPs should be stable. In both cases, MSP temperatures
agree with observations ($T^\infty\lesssim \tmax$). Spin
frequency evolution is almost unaffected by r-modes even
for high-frequency MSPs (see a discussion after equation
\ref{a_gen}). However, the recycling scenario requires that
MSPs should be formed in LMXBs. In the following subsection
we analyse constraints on models (a) and (b) set by
this requirement.

Note that, generally, the resonance temperatures can depend
not only on the microphysics input (i.e., on the equation
of state, critical temperature profile, etc.), but also on
the stellar mass (e.g., the appearance of hyperons in
sufficiently massive NSs can change the form of the
instability window dramatically). Thus, we cannot exclude the fact
that both possibilities (a) and (b) are realized for
different NS masses.

\subsection{Narrow low-temperature stability peaks [model (a)]}
\label{Sec_narrow}

Typical evolution path, which should lead to the formation of
low-temperature MSPs within the model (a), is shown by a
solid line in panel (a) of Fig.\ \ref{Fig_UpLift}.
According to the resonant uplift scenario (see
\citealt{gck14b} for details), at the initial stage of
recycling, an NS is stable and spun up by accretion; its
temperature is determined by the balance between the deep
crustal heating and cooling processes
[$T^\infty=T^\infty_\mathrm{eq}$, where
$T^\infty_\mathrm{eq}$ is the solution to equation
(\ref{Teq})]. If spin up is strong enough, NS can cross the
boundary of the stability region. This leads to excitation
of r-mode and accompanying heating of a star. As a result,
its temperature increases. However, growth of the
temperature is limited by the boundary of the stability
peak, because inside the peak the star is stable and does
not have a heat source to further increase $T^{\infty}$. As
a consequence, subsequent NS evolution takes place along
the boundary of the stability peak. The r-mode amplitude is
determined by the heating power required by the thermal
equilibrium condition (see \citealt{gck14b} for details):
\begin{equation}
W^\mathrm{r}=L_\mathrm{cool}(T^\infty)-W^\mathrm{DCH}
\label{ThermEqul}
\end{equation}

To be associated with the low-temperature stability peak,
located at $T^\infty=T^\infty_0$
[$T^\infty_0$ is the `resonance' temperature; see Fig.\ \ref{Fig_UpLift}(a)],
the equilibrium temperature should be
$T^\infty_\mathrm{eq}\lesssim T^\infty_0$ (in the opposite
case the NS evolution in the unstable region will be
associated with
next stability peak
at $T^\infty=T^\infty_1$). At the same time, to match the
upper limit (\ref{Tmax}) on the MSP temperature, one should
have $T_0^\infty<\tmax=2\times10^7$~K. In this way we
obtain the first important constraint for model (a)
(see equation \ref{Teq}):
\begin{equation}
 \dot M<\dot M_\mathrm{max} \equiv L_\mathrm{cool}(T^\infty_0)/(K\,c^2).
 \label{dotM_LMXB_constr}
\end{equation}

{The second constraint is related to accretion spin-up,
which cannot increase the NS frequency} above the
equilibrium spin frequency
$\Omega_\mathrm{eq}=2\pi\,\nu_\mathrm{eq}$,
when
the net torque acting on the star is zero (for
$\Omega>\Omega_\mathrm{eq}$ the torque is negative). The
exact numerical value of the equilibrium spin rate should
be determined from the accretion theory, but it is still a
subject of debate (see, e.g., \citealt*{lai14_accretion, psb16}
and references therein). The standard model assumes that
the spin equilibrium takes place when the Keplerian disc
corotates with NS at the magnetosphere boundary (see, e.g.,
\citealt{gl79b,wang95} for details)
\begin{equation}
      \Omega_\mathrm{eq}\approx \omega_\mathrm{c}
      \xi^{-3/2}\, \left(G M\right)^{5/7}\,\left(\frac{\sqrt{2}\, \dot M}{\mu^2}\right)^{3/7}.
 \label{nu_std}
\end{equation}
Here $\mu$ is the dipolar magnetic moment of the star;
$\xi\sim 0.5- 1.4$ and $\omega_\mathrm{c}\lesssim 1$ are
the numerical coefficients, which describe radius of the
magnetosphere and spin-up efficiency.

Note that, if accreting NS is attached to the peak and
accretion rate is equal to $\dot M_\mathrm{max}$, the
equilibrium r-mode amplitude (and the corresponding heating
power and breaking torque) are all vanish (see equation
\ref{ThermEqul}). Thus, at such accretion rate the r-mode
instability does not affect the equilibrium frequency
$\Omega_{\rm eq}$. For lower accretion rate, $\Omega_{\rm
eq}$ is smaller and the r-mode braking torque can only
further decrease it. Thus, the maximal equilibrium
frequency can be estimated as
$\Omega^\mathrm{max}_\mathrm{eq}=\Omega_\mathrm{eq}(\dot
M_\mathrm{max})$.

In real LMXBs accretion rate is evolving in time, leading
to the evolution of the equilibrium spin rate (e.g.,
\citealt{Tauris12}). The NS spin frequency $\Omega$ cannot
exceed maximum  equilibrium spin frequency
$\Omega_\mathrm{eq}$, which was achieved during the
evolution.
As far as the accretion rate cannot
exceed $\dot M_\mathrm{max}$
at any moment of the evolution,
the upper bound on the NS spin frequency in LMXB is
$\Omega^\mathrm{max}_\mathrm{eq}$.
Thus, the frequency of
a newly born MSP, $\Omega^\mathrm{max}_\mathrm{MSP}$,
also cannot exceed $\Omega^\mathrm{max}_\mathrm{eq}$.
In fact, as it was shown by \cite{Tauris12},
$\Omega^\mathrm{max}_\mathrm{MSP}$ can be even lower
because MSP can lose more than a half of its rotational
energy during the Roche lobe decoupling phase.
Since at the MSP stage there are no accretion and any spin up torques,
the stellar spin frequency can only decrease,
and, as a result, it should be a subject of our second constraint,
\begin{equation}
    \Omega<\Omega^\mathrm{max}_\mathrm{MSP}=p\, \Omega^\mathrm{max}_\mathrm{eq}(\dot
    M_\mathrm{max},\mu),
    \label{nu_MSP_constr}
\end{equation}
where the numerical parameter $p<1$ describes spin down of
MSP during the Roche lobe decoupling phase.

The third constraint comes from the requirement that the
 spin-down
power, associated with r-modes, $\dot
E_\mathrm{rot}^\mathrm{r}=\Omega \dot J_\mathrm{GR}$,
should not exceed the total spin-down power of MSPs, $\dot
E_\mathrm{rot}^\mathrm{tot}=I\Omega \dot
\Omega^\mathrm{tot}$.
{Here $I$ is NS moment of inertia.}
Consider a NS evolving along the
boundary of stability peak. Then the r-mode heating power
is related to $\dot E_\mathrm{rot}^\mathrm{r}$ by the
formula
\begin{equation}
W^\mathrm r=\frac{1}{3}\left|\dot
E_\mathrm{rot}^\mathrm{r}\right| \label{Wr_spindown},
\end{equation}
which follows from equations (\ref{JGR}) and
(\ref{HeatRate}), applied to the boundary of instability
window with $\tau_\mathrm{eff}=\tau_\mathrm{d}=
\left|\tau_\mathrm{GR}\right|$. In addition, such star
should be in thermal equilibrium, which means that its
luminosity at the peak temperature $T^\infty_0$ should be
equal to $W^\mathrm {r}$, i.e., in view of equation
(\ref{Wr_spindown}),
\begin{equation}
    L_\mathrm{cool}(T^\infty_0)=W^\mathrm r=\frac{\left|\dot E_\mathrm{rot}^\mathrm{r}\right|}{3}
    = \frac{a}{3} \left|\dot E_\mathrm{rot}^\mathrm{tot}\right|
    \le \frac{1}{3}\left|\dot E_\mathrm{rot}^\mathrm{tot}\right|,
    \label{L_MSP}
\end{equation}
where $a=\dot E_\mathrm{rot}^\mathrm{r}/\dot
E_\mathrm{rot}^\mathrm{tot} \le1$ is a (unknown) parameter,
which determines fractional contribution of r-mode spin
down at the present moment. Here we neglect other
mechanisms of internal heating, which can contribute to the
heating power only at a level of $\lesssim 10^{-4}
\left|\dot E_\mathrm{rot}^\mathrm{tot}\right|$ (e.g.,
\citealt{gr10,gkr15}), becoming comparable to the r-mode
heating only for very low values of $a\sim 3\times
10^{-3}$.

In the remaining part of this section we put a lower limit
on $a$ following from the three constraints, discussed
above, and confront this result with observations.

According to \cite*{bgi93,spitkovsky06,tps16}, the
spin-down power associated with magnetic field, $\dot
E_\mathrm{rot}^\mathrm{mag}$, can be estimated as
\begin{equation}
    (1-a)\,\dot E_\mathrm{rot}^\mathrm{tot}=\dot E_\mathrm{rot}^\mathrm{mag}
    \approx -\frac{\mu^2 \Omega^4}{c^3}\left(1+\sin^2\theta\right).
    \label{P_mag}
\end{equation}
Here $\theta$ is the magnetic dipole inclination angle and
we assume that the only torques acting on an MSP are
associated with r-modes and magnetic field
(i.e., $\dot E_\mathrm{rot}^\mathrm{tot}=\dot
E_\mathrm{rot}^\mathrm{r}+\dot
E_\mathrm{rot}^\mathrm{mag}$). One can express $\mu$ from
this equation and substitute it into equation
(\ref{nu_std}) along with $\dot M=\dot M_\mathrm{max}$
[given by equations (\ref{dotM_LMXB_constr}) and
(\ref{L_MSP})], to get $\Omega^\mathrm{max}_\mathrm{eq}$.
The result is
\begin{equation}
  \Omega^\mathrm{max}_\mathrm{eq}\approx
      \omega_c\, \xi^{-3/2}\, \left(G M\right)^{5/7}\,
      \left(\frac{\sqrt{2}\,a\, \Omega^4\, \left(1+\sin^2\theta\right)
      }{
      3\,K\,(1-a)\, c^5}\right)^{3/7}.
        \label{nu_std_fin}
\end{equation}
Note that $\dot E_\mathrm{rot}^\mathrm{tot}$ is cancelled out
in this final equation. It holds true for any accretion
model (in particular, for the recent model of
\citealt*{psb16}), provided that the equilibrium frequency
depends on $\dot M$ and $\mu$ only in combination $\dot
M/\mu^2$.

Equation (\ref{nu_std_fin}) combined with constraint
(\ref{nu_MSP_constr}) allows one to put a lower limit on
the parameter $a$,
\begin{equation}
        a>a_\mathrm{min}\approx
        \left[1
        +0.8 (p\,\omega_c)^{7/3} \nu_{500}^{5/3} M_{1.4}^{5/3}
       \left(1+\sin^2\theta\right)\,\xi^{-7/2}\right]^{-1}.
       \label{a_min}
\end{equation}
%

\begin{figure}
        \includegraphics[width=\columnwidth]{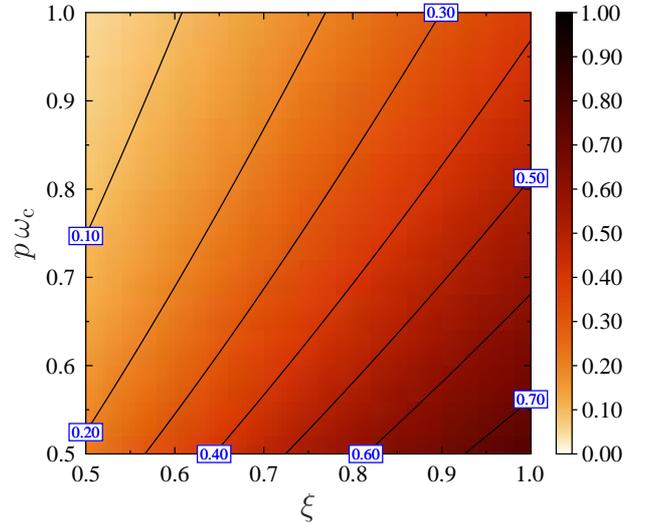}
    \caption{Example of a contour plot showing the minimum possible ratio
        $a_{\rm min}$
        of the r-mode spin-down power to the total spin-down power
    in $(\xi,\,p\omega_c)$ plane.
    To plot the figure we choose
    $\nu=500$~Hz, $\sin\theta=1$, and $M=1.4 M_\odot$.
    }
    \label{Fig_a_min}
\end{figure}

The isolines of $a_\mathrm{min}$ are shown in Fig.\
\ref{Fig_a_min}. MSP with  $M=1.4 M_\odot$, $\nu=500$~Hz,
and $\sin\theta=1$ is taken as an example. Only in the most
efficient spin-up scenario ($\xi\approx 0.5$,
$p\,\omega_c\approx 1 $) $a_\mathrm{min}$ can be as low as
0.05. If at least one of the parameters of the accretion
model is not so optimistic, $a_\mathrm{min}$ is much
larger. The larger value of $a_{\rm min}$ is favoured for
many reasons. In particular, \cite*{tlk12} argue that
$\omega_c<1$ and $\xi>1$ for at least some of MSPs; the
most efficient model ($p=1$) does not account for spin-down
both during the Roche lobe decoupling phase
(\citealt{Tauris12}) and at the MSP stage; recent model by
\cite{psb16} suggests that $\omega_c$ should be very small
for small $\xi$; finally, smaller angle $\theta$ leads to
higher $a_\mathrm{min}$. Furthermore, to obtain an upper
limit for the spin frequency (\ref{nu_MSP_constr}) we
assume  accretion rate to be equal to its upper limit,
$\dot M=\dot M_\mathrm{max}$. If real $\dot M$ is lower,
the equilibrium spin frequency is lower too, leading to a
stronger lower bound on $a$. All in all, we believe that a
realistic value of $a_\mathrm{min}$ should exceed $0.05$.

Let us confront the constraint (\ref{a_min}) with
observations. Unfortunately, direct detection of
gravitational wave signal from r-modes is impossible,
because (at least, for high frequency MSPs, $\nu\ge 300$)
it is below the sensitivity level of advanced LIGO due to
very small r-mode amplitudes allowed in MSPs (see section
\ref{Sec_tiny} and, e.g., \citealt{as15_oscMSP}). Another
almost straightforward procedure to measure $a$ is
associated with the so called $\alpha$-oscillations --
modulations of r-mode amplitude, which are unavoidable for
MSPs evolving along
 a narrow
stability peak
(\citealt{kgc16}). This modulation leads to `anti-glitches'
-- sudden drops of MSP frequency on a time-scale of
hours-months. Such features have not been found yet in MSP
timing, thus observations can be used to put an upper limit
on $a$. Indeed, in the limiting case $a\approx 1$, almost
all MSP spin-down should be associated with `anti-glitch'
events, which clearly contradicts observations. At the same
time, our preliminary results indicate that $a=0.05$ can
agree with timing residuals for PSR B1937+21, reported in
\cite{PSR1937new}. We leave the problem of accurate
determination of an upper limit for $a$ to subsequent
studies.

Constraint (\ref{a_min}) can also be checked indirectly.
According to equation (\ref{Wr_spindown}), r-mode
contribution to the spin down unavoidably leads to NS
heating. Thus, equation (\ref{a_min}) determines the
minimum heating power for MSP evolving along a narrow
low-temperature stability peak,
\begin{equation}
 W_\mathrm{min}^\mathrm{r}=\frac{a_\mathrm{min}}{3}\,
 \left |\dot E_\mathrm{rot}^\mathrm{tot}\right|. \label{Q_min}
\end{equation}
This heating can be too strong to be balanced by cooling at
low temperature.
For example, taking the moment of inertia
$I=10^{45}$~g\,cm$^2$, PSR B1937+21 has $\left|\dot
E_\mathrm{rot}^\mathrm{tot}\right|\sim
10^{36}$~erg\,s$^{-1}$. Thus, for the lowest possible value
of $a_\mathrm{min}=0.05$, the minimal r-mode heating power
is $W_\mathrm{min}^\mathrm{r} \sim 2\times
10^{34}$~erg\,s$^{-1}$.
Even assuming that fully unsuppressed hyperon URCA is open
in the central region with the radius $R_\mathrm{D}=3$\,km,
this heating power is sufficient to heat the star up to a
temperature $T^\infty\approx 3\times 10^7$~K [see equation
(\ref{L_DURCA})], which exceeds the upper limit
(\ref{Tmax}) by a factor of 1.5. As noted above, if
accretion spin-up is not extremely efficient,
$a_\mathrm{min}$ should be much larger, substantially
complicating the explanation of MSPs with $\left|\dot
E_\mathrm{rot}^\mathrm{tot}\right|\gtrsim 10^{35}$~erg
s$^{-1}$. Thus, the model with narrow low-temperature
resonance peaks should be rejected (at least) for pulsars
with high spin-down power. But what about the model (b)
with broad low-temperature stability peaks?

\subsection{Broad low-temperature stability peaks [model (b)]}
\label{Sec_broad}

Typical evolution path in case of broad
low-temperature peaks [model (b)] is shown in panel (b)
of Fig.\ \ref{Fig_UpLift}.
During the LMXB stage an NS evolves inside the
stability peak.
Its temperature is determined
by the deep crustal heating (see equation \ref{Teq}) and
the spin frequency is controlled by accretion. After
Roche-lobe decoupling [filled circle in panel (b) of Fig.\
\ref{Fig_UpLift}]
accretion stops, and the NS becomes an MSP, which cools
down by photon and neutrino emission. If its frequency is
high enough
[$\gtrsim 594$~Hz for the example shown in Fig.\
\ref{Fig_UpLift}(b)], cooling stops at the left boundary of
the stability peak and the r-mode excites up to an
amplitude required to keep NS in thermal equilibrium [see
equation (\ref{ThermEqul}) with $W^{\rm DCH}=0$]. In the
opposite case, MSP stays stable (r-mode is not excited) and
evolves to the left until the thermal equilibrium with
other (less effective) heating processes (see section
\ref{Sec_timing}) is established.

In order to explain observations of MSPs  model (b)
should satisfy two main requirements:
\begin{description}
\item{(A)} Temperature of the \underline{left} slope of
the low-temperature stability peak should not exceed the
observational constraint (\ref{Tmax}) in the whole MSP
frequency range.
\item{(B)} Accretion should be able to spin the star up to
high frequencies {\it within} the low-temperature peak [in
the opposite case, immediately after an
 NS reaches the high-temperature
boundary of the peak, the r-modes will be excited, heating
the star up to the next (high-temperature) stability peak,
so that the subsequent evolution will be associated with
that peak].
\end{description}

The constraint (\ref{a_min}) is unjustified in case of
broad low-temperature resonances because the cooling
luminosity decreases significantly during cooling of a
newly born MSP inside the stability peak. As a result, the
contribution of r-modes to the spin down of MSPs can be
very small.

Both requirements (A) and (B) can be satisfied by adjusting
the parameters of the low-temperature stability peak. For
example, in Fig.\ \ref{Fig_UpLift}(b) the requirement (A)
holds true up to $\nu<800$~Hz.
The requirement (B) can be satisfied by adjusting the high-temperature slope
    of stability peak, as it was done in Fig.\ \ref{Fig_UpLift}(b).
    Indeed, the fact that a number
    of high-frequency accreting NSs (which are progenitors of
    MSPs according to the recycling scenario) are contained
    inside the broad stability peak
    {shown in \ \ref{Fig_UpLift}(b)} can serve as an indirect
    evidence that the requirement (B) is fulfilled
    {for this model}.

We can also check the consistency of a broad-peak model
with requirement (B) for certain accretion model. For
example, the standard accretion model predicts the
following minimum accretion rate, required to spin up an
(r-mode stable) MSP with the observed $\nu$ and $\dot \nu$,
\begin{equation}
  \dot  M_\mathrm{min}=
  \frac{c^3\xi^{7/2}}{2^{1/2} G^{5/3}p^{7/3} \omega_c^{7/3} \left(1+\sin^2\theta\right)}
    \frac{I\dot \Omega}{ M^{5/3} \Omega^{2/3}}
\end{equation}
[see, e.g., \citealt{tlk12}; the same result can also be
derived from equations (\ref{nu_std}),
(\ref{nu_MSP_constr}), and (\ref{P_mag})]. This accretion
rate corresponds to the minimal power of deep crustal
heating,
\begin{eqnarray}
 W^\mathrm{DCH}_\mathrm{min}&=& c^2 K \dot  M_\mathrm{min}
\approx
 3.6\times 10^{33}\mbox{erg\,s}^{-1} \frac{\dot \nu_{14}\,I_{45}
 }
 {\nu_{500}^{2/3}\, M_{1.4}^{5/3}},
 \label{Q_min_DCH}
\end{eqnarray}
where in the last equality we assume $\xi=0.5$,
$p=\omega_c=\sin(\theta)=1$, $K=10^{-3}$,  $\dot
\nu_{14}=\dot \nu/10^{-14}\mbox{\,Hz\,s}^{-1}$, and
$I_{45}=I/10^{45}$~g\,cm$^2$. For a (stable) NS inside the
stability peak this heating should be compensated by
cooling, providing a constraint on the \textit{minimal} NS
temperature at the LMXB stage, $T^\infty\gtrsim
T^\infty_\mathrm{eq,\,min}$, where
$T^\infty_\mathrm{eq,\,min}$ is found from
\begin{equation}
L_\mathrm{cool}(T^\infty_\mathrm{eq,\,min})=W^\mathrm{DCH}_\mathrm{min}.
\label{Tmin_LMXB}
\end{equation}
For example, the strongest (among the known MSPs)
$W^\mathrm{DCH}_\mathrm{min}\approx 1.4\times
10^{34}$~erg\,s$^{-1}$ corresponds to PSR B1937+21
(assuming $I_{45}=1$). It can be compensated by
unsuppressed hyperon direct URCA (assuming
$R_\mathrm{D}=3$\,km) at
$T^\infty_\mathrm{eq,\,min}\approx 3.1\times
10^{7}~\mathrm{K}$. 
Thus, the temperature of this NS at LMXB stage should not
be less than this value. At $T^\infty= 3.1\times 10^{7}$~K
the high frequency boundary of the broad stability peak in
Fig.\ \ref{Fig_UpLift}(b) corresponds to $\nu\approx
750$~Hz, thus PSR B1937+21 (with $\nu\approx 642$~Hz)
could, in principle, be spun up  by accretion on the LMXB
stage, without leaving out this stability peak.

To explain NSs observed in LMXBs with higher temperatures,
our model (b) requires also high-temperature stability
peaks. As it is argued in \cite{cgk14}, NSs associated with
such peaks evolve not to MSPs, but to so called `HOFNARs'.
{These NSs maintain high temperature due to r-mode
instability even after the end of accretion epoch in LMXBs
and should be observed in X-rays because of their high
surface temperatures, $T^\infty_\mathrm{eff}\sim 10^6$\,K.
Some of the candidates to quiescent LMXB systems (qLMXBs)
-- X-ray sources, whose spectra
are interpreted as
thermal emission from
the whole NS surface, but accretion episodes have never been detected
(see, e.g., \citealt{Guillot_etal11}) -- can, in fact, belong to that new class.
As discussed in \cite{cgk14} HOFNARs can have very
low magnetic fields due to Ohmic dissipation and thus do
not
reveal themselves
as radio pulsars.
Non-detection
(to our best knowledge)
of
radio emission from qLMXB candidates
($=$HOFNARs candidates) supports this statement.}
 The lower bound
on the r-mode contribution to the spin-down (equation
\ref{a_min}) can also be applied to HOFNARs, but it is not
restricting, because their spin-down can, in fact, be
generated purely by r-modes.

$\alpha$-oscillations can be suppressed for NS evolving
along the broad stability peaks (\citealt{kgc16}). Thus,
absence of `anti-glitches' would not allow one to draw any
conclusion about validity of the model with broad stability
peak. However, if anti-glitches were observed, it would
provide strong constraints on the parameters of the
stability peaks.

\section{General constraints on the r-mode instability window}
\label{Sec_gen}

{
\begin{figure}
        \includegraphics[width=\columnwidth]{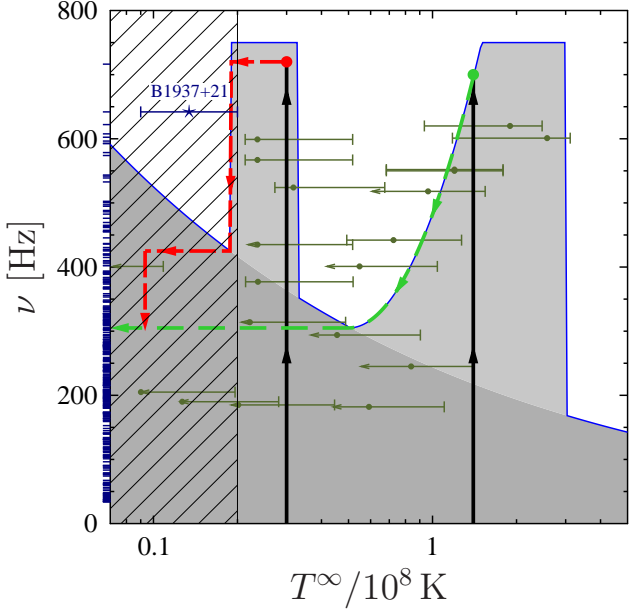}
    \caption{
        {An example of a minimally constrained r-mode instability
    window. The evolution tracks, leading to the formation of MSPs and
    HORNARs are shown by thick lines with arrows
    (solid
    lines correspond to evolution at the LMXB stage, dashed
    lines --- after it). Regions of enhanced stability (in comparison
    to the minimal model of section \ref{Sec_minmod}) are
    shown by lighter shading. Other notations are the same,
    as in Fig.\ \ref{Fig_MinMod}.
    }}
    \label{Fig_minwin}
\end{figure}
}

 The constraints on the resonance uplift scenario obtained
in section \ref{Sec_reson} appeal only to the shape of the
instability window, but not to the actual physics which
determines that shape. Thus, our results are equally
applicable to arbitrary r-mode instability model with
similar morphology of the stability region
(a set of humps of enhanced stability located at certain temperatures%
{, see Fig.\ \ref{Fig_minwin}}).%
\footnote{More complicated stability windows, e.g., like
those suggested in \cite{gck14b} (see Appendix D there),
cannot also be excluded.}
They can be formulated as follows
{(an example of the instability window, satisfying the
constraints below, is shown in Fig.\ \ref{Fig_minwin})}
:
%
\begin{description}
\item[(I)]
To explain formation of MSPs in agreement with the upper
limit (\ref{Tmax}) on MSP temperature, the r-mode
instability should be suppressed in the whole MSP frequency
range ($\nu\lesssim 750$~Hz) at least for temperatures
$T^\infty_\mathrm{max}\lesssim T^\infty\lesssim
T^\infty_\mathrm{eq,\,min}$ [cross-hatched region in Fig.\
\ref{Fig_UpLift}(b)]. Here $T^\infty_\mathrm{eq,\,min}$ is
given by equation (\ref{Tmin_LMXB}).
$T^\infty_\mathrm{eq,\,min}$ is specific for each MSP, but
the strongest bound (the largest
$T^\infty_\mathrm{eq,\,min}$) comes from PSR B1937+21. In
case of the standard accretion model and the direct URCA
emissivity  (see equation \ref{L_DURCA}) it corresponds to
$T^\infty_\mathrm{eq,\,min}\sim 3.1\times 10^7$~K (assuming
$R_\mathrm{D}=3$\,km) and becomes higher if cooling and/or
accretion spin up is less efficient.

\item[(II)]
The r-mode instability can be unsuppressed at low
temperatures ($T^\infty<T^\infty_\mathrm{max}$) and, if it
is the case, it can, in principle, be confirmed by
observations. Namely, if minimum of the instability curve
for $T^\infty<T^\infty_\mathrm{max}$ takes place at a point
$(T^\infty_\mathrm{min},\ \nu_\mathrm{min})$ (e.g., in
Fig.\ \ref{Fig_UpLift}(b) $T^\infty_\mathrm{min}\approx
10^7$~K and $\nu_\mathrm{min}\approx594$~Hz), {\it all}
MSPs with $\nu>\nu_\mathrm{min}$ should have
$T^\infty>T^\infty_\mathrm{min}$ and can be affected by
anti-glitches, associated with $\alpha$-oscillations (if
they are not suppressed).

\item[(III)] Strictly speaking, current observations do not require to stabilize all NSs
observed in LMXBs by one hump -- the r-mode instability can
be almost unsuppressed in the region  $7\times
10^7\mathrm{~K}\lesssim T^\infty\lesssim 10^8$~K [as in
Figs.\ \ref{Fig_UpLift}(b) and \ref{Fig_minwin}]. However, at larger temperatures
instability should be suppressed by additional hump (may be
narrow) to explain the hottest NSs in LMXBs with
$T^\infty\sim 10^8$~K (e.g., 4U 1608-522). Such stars
should evolve
 not to MSPs, but to
HOFNARs, as suggested by \cite{cgk14} {(see high
temperature stability hump in Fig.\ \ref{Fig_minwin} and
correspondent evolution track)}.

\end{description}

At the end let us note that the constraint (I) can be used
to quickly check
{various r-mode instability models  for consistency with
fiducial MSP
temperature upper bound  (\ref{Tmax}), which assumes
accreted thermally insulating envelope (see section
\ref{Sec_Xray_MSP} for details).}
{ For example, some of the models discussed in the recent
review by \cite{haskell15} disagree with this constraint.
Namely, the model with strong mutual friction
and `strong' superfluidity (left panel of Fig.\ 3 in that
paper) predicts that all MSPs with $\nu\gtrsim550$\,Hz
should have internal temperature $T\gtrsim 3\times
10^7$\,K.
It can agree with (\ref{Teffmax}) only for the layer
of accreted material $\Delta M\lesssim 10^{-12}\,M_\odot$
(assuming a canonical NS).
The strange star model
(Fig.\ 4 of \citealt{haskell15}) predicts even larger temperatures,
requiring
lower
$\Delta M$.
}

\section{Conclusions}
\label{Sec_Consl}

We put new constraints on the shape of the r-mode
instability window. They follow from the recycling scenario
and observations of (a) NSs in LMXBs (section
\ref{Sec_Xray}), (b) MSP timing (section \ref{Sec_timing})
and (c) \textit{UV} and X-ray observations of MSPs (section
\ref{Sec_Xray_MSP}). Analysing the observational sets (a)
and (b), which have already been used to
put an upper bound on
the r-mode instability by
several authors, we emphasize the importance of two points:
(i) any successful model should not only suppress the
instability for NSs observed in LMXBs, but also explain
their temperatures (this is especially important for models
based on an enhanced bulk viscosity, see section
\ref{Sec_bulk}); (ii)
{internal heating mechanisms [superfluid vortex creep
 (\citealt{Alpar_etal84}), rotochemical heating
(\citealt{Reisenegger95}), and rotation-induced deep
crustal heating (\citealt*{gkr15})] are  strong
enough  to make PSR B1937+21 unstable within the minimal
r-mode instability model, providing thus new evidence of
additional dissipation of r-modes at low temperatures. }
One more
constraint comes from the observational set (c),
which allows us to put an upper limit on the internal MSP
temperature, $T^\infty\lesssim T_{\rm max}^{\infty} \approx
2\times 10^7$~K, limiting thus the maximum possible r-mode
amplitude in MSPs
(section \ref{Sec_tiny}). These results qualitatively agree
with similar conclusions of \cite{Schwenzer_etal_Xray}.
Combined with the recycling scenario of MSP formation, the
temperature upper limit (\ref{Tmax}) allows us to conclude
that the r-mode instability should be suppressed in the
whole MSP frequency range ($\nu\lesssim 750$~Hz) at
temperatures $2\times 10^7\mathrm{~K}\lesssim
T^\infty\lesssim 3 \times 10^7$~K (see
Fig.
 \ref{Fig_minwin} and section \ref{Sec_gen} for
details). This
condition can be used to quickly
check for consistency various instability windows available
in the literature. In particular, it can be fulfilled in
the resonance uplift scenario suggested by
\cite{gck14a,gck14b} (see section \ref{Sec_reson} for a
detailed analysis).

It is worth noting, that constraining the surface
temperature of the fastest MSPs by detailed analysis of
X-ray and \textit{UV} observational data can be a very interesting
task. It is very likely that such analysis can put much
more stringent upper limits on the surface temperatures of
individual sources, than our general upper limit
$T_\mathrm{eff}^\infty\lesssim 6\times 10^5$~K (see section
\ref{Sec_Xray_MSP}, and \citealt{Schwenzer_etal_Xray}).
This can make the r-mode instability even more tightly
constrained. In the opposite case, if one observes an MSP
with higher temperature, it will be even more important,
because it will present an almost direct evidence of r-mode
instability in such pulsar: The thermal emission from the
surface $L_\mathrm{cool}^\mathrm{ph}\gtrsim
10^{32}$~erg\,s$^{-1}$ should have an energy source, and the
r-mode instability seems to be the most natural `energy
supplier' in the absence of accretion.%
\footnote{Other internal heating mechanisms have efficiency
$W^\mathrm{i}\lesssim 10^{-4}\left |\dot
E_\mathrm{rot}^\mathrm{tot}\right|$ (see
\citealt{gr10,gkr15}), producing $W^\mathrm{i}\lesssim
10^{32}$~erg\,s$^{-1}$ even for MSPs with the highest spin
down power, $\left|\dot
E_\mathrm{rot}^\mathrm{tot}\right|\sim
10^{36}$\,erg\,s$^{-1}$.}

\section*{Acknowledgements}
We are very grateful to Dima Zyuzin for suggesting us the
paper by \cite{hnk13} and to an anonymous referee for
useful comments. This study was supported by the
Russian Science Foundation (Grant No. 14-12-00316).

\label{lastpage}

\end{document}